\documentclass[aps,prl,twocolumn,preprintnumbers,amsmath,amssymb, superscriptaddress, bibnotes,longbibliography]{revtex4}
\usepackage[utf8]{inputenc}
\usepackage{graphicx}
\usepackage{dcolumn}
\usepackage{bm}
\usepackage[colorlinks,linkcolor=blue,anchorcolor=blue,citecolor=blue,urlcolor=blue,filecolor=blue,menucolor=blue,runcolor=blue]{hyperref}
\usepackage{ulem}
\usepackage{threeparttable}

\makeatletter
\@ifundefined{textcolor}{}
{%
	\definecolor{BLACK}{gray}{0}
	\definecolor{WHITE}{gray}{1}
	\definecolor{RED}{rgb}{1,0,0}
	\definecolor{GREEN}{rgb}{0,1,0}
	\definecolor{BLUE}{rgb}{0,0,1}
	\definecolor{CYAN}{cmyk}{1,0,0,0}
	\definecolor{MAGENTA}{cmyk}{0,1,0,0}
	\definecolor{YELLOW}{cmyk}{0,0,1,0}
}

\usepackage{color}

\newcommand{\beq}{\begin{eqnarray}}
\newcommand{\eeq}{\end{eqnarray}}

\newcommand{\CCS}{{CsCr$_3$Sb$_5$}}
\newcommand{\AVS}{{$A$V$_3$Sb$_5$}}

\makeatother
\usepackage[english]{babel}

\begin{document}

\title{Antiferromagnetic Dimers in the Parent Phase of a Correlated Kagome Superconductor}

\author{Yifan Wang}
\thanks{These authors contributed equally.}
\affiliation{Center for Correlated Matter and School of Physics, Zhejiang University, Hangzhou 310058, China}

\author{Chenchao Xu}
\thanks{These authors contributed equally.}
\affiliation{School of Physics, Hangzhou Normal University, 310036 Hangzhou, China}

\author{Yi Liu}
\affiliation{School of Physics, Zhejiang University, Hangzhou 310058, China}
\affiliation{Department of Applied Physics, Key Laboratory of Quantum Precision Measurement of Zhejiang Province, Zhejiang University of Technology, Hangzhou, China}

\author{Jinke Bao}
\affiliation{School of Physics, Hangzhou Normal University, 310036 Hangzhou, China}

\author{Jiayu Guo}
\affiliation{Center for Correlated Matter and School of Physics, Zhejiang University, Hangzhou 310058, China}

\author{Xiaoran Yang}
\affiliation{Center for Correlated Matter and School of Physics, Zhejiang University, Hangzhou 310058, China}

\author{ Yuiga Nakamura}
\affiliation{Diffraction and Scattering Division, Japan Synchrotron Radiation Research Institute, SPring-8, Sayo, Hyogo 679-5198, Japan}

\author{Hiroshi Fukui}
\affiliation{Precision Spectroscopy Division, Japan Synchrotron Research Institute, SPring-8, Kouto, Sayo, Hyogo 679-5198, Japan}

\author{Taishun Manjo}
\affiliation{Precision Spectroscopy Division, Japan Synchrotron Research Institute, SPring-8, Kouto, Sayo, Hyogo 679-5198, Japan}

\author{Daisuke Ishikawa}
\affiliation{Materials Dynamics Laboratory, RIKEN SPring-8 Center, Sayo, Hyogo 679-5148, Japan}
\affiliation{Precision Spectroscopy Division, Japan Synchrotron Research Institute, SPring-8, Kouto, Sayo, Hyogo 679-5198, Japan}

\author{Alfred Q. R. Baron}
\affiliation{Materials Dynamics Laboratory, RIKEN SPring-8 Center, Sayo, Hyogo 679-5148, Japan}
\affiliation{Precision Spectroscopy Division, Japan Synchrotron Research Institute, SPring-8, Kouto, Sayo, Hyogo 679-5198, Japan}

\author{Saizheng Cao}
\affiliation{Center for Correlated Matter and School of Physics, Zhejiang University, Hangzhou 310058, China}

\author{Rui Li}
\affiliation{Center for Correlated Matter and School of Physics, Zhejiang University, Hangzhou 310058, China}

\author{Zilong Li}
\affiliation{Center for Correlated Matter and School of Physics, Zhejiang University, Hangzhou 310058, China}

\author{Yanan Zhang}
\affiliation{Center for Correlated Matter and School of Physics, Zhejiang University, Hangzhou 310058, China}

\author{Ruihan Chen}
\affiliation{Center for Correlated Matter and School of Physics, Zhejiang University, Hangzhou 310058, China}








\author{Ming Shi}
\affiliation{Center for Correlated Matter and School of Physics, Zhejiang University, Hangzhou 310058, China}
\affiliation{Institute for Advanced Study in Physics, Zhejiang University, Hangzhou 310058, China}

\author{Huiqiu Yuan}
\affiliation{Center for Correlated Matter and School of Physics, Zhejiang University, Hangzhou 310058, China}
\affiliation{State Key Laboratory of Silicon and Advanced Semiconductor Materials, Zhejiang University, Hangzhou 310058, China}
\affiliation{Institute of Fundamental and Transdisciplinary Research, Zhejiang University, Hangzhou 310058, China}

\author{Guanghan Cao}
\affiliation{School of Physics, Zhejiang University, Hangzhou 310058, China}
\affiliation{State Key Laboratory of Silicon and Advanced Semiconductor Materials, Zhejiang University, Hangzhou 310058, China}
\affiliation{Institute of Fundamental and Transdisciplinary Research, Zhejiang University, Hangzhou 310058, China}

\author{Chao Cao}
\email{ccao@zju.edu.cn}
\affiliation{Center for Correlated Matter and School of Physics, Zhejiang University, Hangzhou 310058, China}

\author{Yu Song}
\email{yusong\_phys@zju.edu.cn}
\affiliation{Center for Correlated Matter and School of Physics, Zhejiang University, Hangzhou 310058, China}

\begin{abstract}
Kagome metals are prone to charge-density wave (CDW), magnetic, and superconducting phases, with their flat electronic band conducive for correlated physics. In contrast to the weakly correlated {\AVS} ($A$ = K, Rb, Cs) kagome metals with a $2\times2$ CDW, {\CCS} is a correlated metal with a flat band close to the Fermi level, and exhibits a $4\times1$ CDW intertwined with magnetic order. Under pressure, the intertwined orders are suppressed and give way to a dome of superconductivity that emerges from a non-Fermi liquid normal state. Here, we solve the crystal structure of the $4\times 1$ CDW state in {\CCS}, and show it consists of Cr dimers separated by Cr chains. First-principles calculations show the dominant exchange interaction is antiferromagnetic within the dimers, while the intra-chain and dimer-chain couplings are much weaker. The CDW transition of {\CCS} is found to be more strongly first-order than those in {\AVS}, without significant soft phonons or diffuse scattering above the CDW transition temperature. These findings suggest that fluctuating antiferromagnetic dimers may play a major role in the electron pairing of superconducting {\CCS}. 
\end{abstract}

\maketitle

\section{Introduction}
A common feature of unconventional superconductivity (SC) in correlated materials is proximity to magnetic instabilities \cite{Scalapino2012,Keimer2015,Dai2015}, which can be intertwined with a symmetry-breaking electronic order, leading to charge or structural modulations \cite{Fradkin2015,Fernandes2019,Bhmer2022}. Upon appropriate tuning, magnetic or intertwined orders may be destabilized at a quantum critical point (QCP), where the proliferation of corresponding fluctuations may mediate a dome of unconventional SC \cite{Uemura2009,White2015,Hu2024}. In this context, an understanding of the ordered phase provides critical information for fluctuations at the QCP, and the SC that they potentially drive.

The kagome lattice hosts van Hove singularities and flat bands, which may lead to unconventional SC and charge-density waves (CDWs) \cite{Mielke1991,Yu2012,Kiesel2012,Wang2013,Kiesel2013}. The discovery of SC coexisting with a CDW in the {\AVS} ($A$ = K, Rb, Cs) kagome metals \cite{Ortiz2020,Wilson2024} led to proposals of unconventional pairing \cite{Zhao2024,Guguchia2023,Deng2024}, while there is also compelling evidence for conventional SC \cite{Duan2021,Mine2024,Kaczmarek2025,Zhou2024}. Furthermore, the CDW in {\AVS} collapses via first-order-like transitions under pressure, and SC persists well beyond the CDW phase boundary \cite{Yu2021,Chen2021,Du2021}, rather than forming a singular dome anchored around a QCP. The nonmagnetic and weakly correlated nature of {\AVS} \cite{Ortiz2020,Zhao2021}, as well as the absence of QCPs in their pressure-temperature phase diagrams, differentiate these compounds from the paradigm of SC in strongly correlated metals \cite{Uemura2009,White2015,Hu2024}.

\begin{figure*}[t]
    \centering
    \includegraphics[width=1\linewidth]{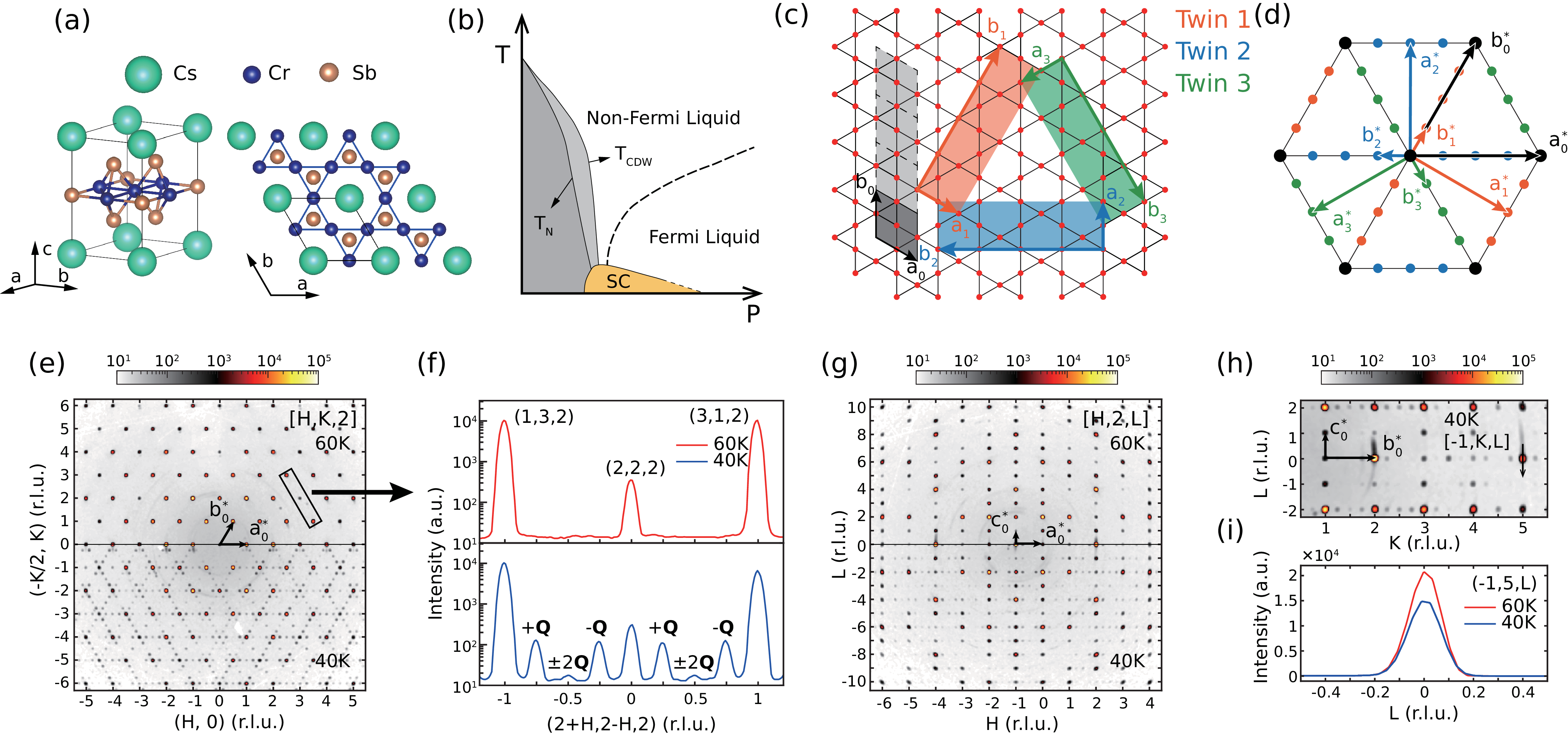}
    \caption{{\bf Physical properties and single crystal XRD data of {\CCS}.} (a) Schematic crystal structure of {\CCS} above $T_{\rm CDW}$. (b) Schematic pressure-temperature phase diagram of {\CCS}, with a dome of SC emerging at the border of intertwined CDW and magnetic orders, from a non-Fermi liquid normal state. (c) The CDW leads to a $4a_0\times a_0$ expansion of the unit cell in the $ab$-plane (shaded gray parallelogram enclosed by dashed lines), which could also be described using a $2\sqrt{3}a_0\times a_0$ orthorhombic cell (shaded orange rectangle). There are two other twin domains related by 120$^\circ$ and 240$^\circ$ rotations (green and blue rectangles). (d) Reciprocal lattice of {\CCS} in the CDW phase, with integer positions for each twin domain indicated by dots with the corresponding color, and the main Bragg peak positions indicated by larger black dots. (e) XRD images of {\CCS} in the $[HK2]$ plane, with the upper half at 60~K (above $T_{\rm CDW}$) and the lower half at 40~K (below $T_{\rm CDW}$). (f) A representative cut from (e) at 40~K and 60~K, with clear CDW peaks at 40~K. (g) $[H2L]$ maps at 60~K and 40~K. (h) $[H1L]$ map (upper panel) and (i)  $(41L)$ cut for 40~K and 60~K, showing no clear splitting of the ${\bf Q}=(4,1,0)$ peak below $T_{\rm CDW}$. It should be noted that $(4,1,0)$ is related to $(-1,5,0)$ via a 60$^{\circ}$ rotation. }   \label{fig:overall}
\end{figure*}

Recently, a correlated kagome metal {\CCS} with the same high-temperature structure as {\AVS} [Fig.~\ref{fig:overall}(a)] was discovered. Compared to {\AVS} that shows a $2\times2$ in-plane charge modulation without magnetic order, {\CCS} exhibits a $4\times 1$ in-plane charge modulation coupled with antiferromagnetic (AFM) order below $T_{\rm CDW}=T_{\rm N}\approx 55$~K. Specific heat measurements reveal an enhanced Sommerfeld coefficient $\gamma\approx100$~mJ$\cdot$K$^{-1}\cdot$mol$^{-1}$, indicative of significant electronic correlations \cite{Liu2024}. Angle-resolved photoemission spectroscopy (ARPES) measurements reveal a flat band slightly below the Fermi level without band folding across $T_{\rm CDW}$ \cite{Peng2024,Li2025,Wang2025}, and resonant inelastic X-ray scattering uncovered robust dipersionless magnetic excitations above and below $T_{\rm CDW}$ \cite{Wang2025}. Scanning tunneling microscopy (STM) measurements found a $4\times \sqrt{3}$ CDW coexistent with the $4\times1$ CDW on the surface of {\CCS} \cite{Cheng2026,Li2025a,Huang2025}. Under pressure, the coupled CDW and magnetic transitions split in temperature and are gradually suppressed, both disappearing around a putative QCP at 4~GPa, giving way to a dome of SC that emerges from a non-Fermi-liquid normal state [Fig.~\ref{fig:overall}(b)] \cite{Liu2024}. Such a pressure-temperature phase diagram of {\CCS} contrasts with those of {\AVS}, where Fermi-liquid behavior is retained across a first-order collapse of the CDW, and SC persists well beyond the full suppression of the CDW \cite{Yu2021,Chen2021,Du2021}. Therefore, while structurally similar to {\AVS}, {\CCS} exhibits physical properties and a pressure-temperature phase diagram akin to correlated superconductors.

A central unanswered question in {\CCS} is the nature of its intertwined charge and magnetic orders, whose dynamic variants may persist under pressure and account for both SC and the non-Fermi-liquid normal state. Given the intertwined nature of charge and magnetic orders in {\CCS}, investigations of its magnetic order requires knowledge of its CDW. Density functional theory (DFT) calculations enumerated collinear magnetic ground states, and found a $4\times2$ charge modulation to be lowest in energy, with the corresponding $\mathbf{q}=0$ AFM order categorized as altermagnetic \cite{Xu2025}. The discrepancy between DFT (a $4\times2$ CDW state lowest in energy) \cite{Xu2025} and experiments (a $4\times1$ CDW in the ground state) \cite{Liu2024} is reflective of the correlated nature of {\CCS} and its frustrated energy landscape of competing phases. These results underscore the need for an experimental determination of the CDW structure in {\CCS}, which can then be used to model its magnetic ground state and magnetic interactions. 

Here, we experimentally determine the crystal structure of {\CCS} in its CDW state. The $4\times 1$ modulation arises from the formation of Cr dimers separated by Cr chains. The Cr-Cr bond length is $\approx2.56$~{\AA} for the dimers, whereas other Cr-Cr bonds average to $\approx2.78$~{\AA}. Based on the experimentally determined crystal structure, we performed DFT calculations to determine the magnetic ground state and associated magnetic interactions, which reveal a ${\bf q}=0$ altermagnetic state with a dominant AFM coupling within the dimer. Our findings form the basis for understanding the physics of {\CCS}, particularly intriguing of which is the connection between AFM dimers in the parent phase, and Cooper pairs in the superconducting phase. 

\section{Results}
\subsection{Structure of the CDW state}

{\CCS} has the same high-temperature $P6/mmm$ structure as {\AVS}, with two-dimensional networks of Cr$_3$Sb$_5$ separated by Cs layers, and magnetic Cr atoms forming kagome lattices [Fig.~\ref{fig:overall}(a)] \cite{Liu2024}. Previous single crystal X-ray diffraction (XRD) measurements showed the CDW of {\CCS} is modulated by ${\bf q}_{\rm CDW}=(1/4,0,0)$, and thus has a $4a_0\times a_0$ unit cell in real space [dashed line in Fig.~\ref{fig:overall}(c)]. Equivalently, the CDW modulation can be described using an orthogonal $2\sqrt{3}a_0\times a_0$ unit cell that has the same volume, with three twin domains [colored rectangles in Fig.~\ref{fig:overall}(c)]. To distinguish between momenta indexed for the CDW phase and the high-temperature phase, a subscript $i$ ($i=1,2,3$) is used to denote a vector in reciprocal lattice units of the CDW twin $i$, whereas vectors without subscripts index momenta for the high temperature phase. In this notation, $(H,K)_i$ ($i=1,2,3$) and $(H,K)$ are related via:

\begin{equation*}
\begin{pmatrix}
H\\
K\\
\end{pmatrix}_1=
\begin{pmatrix}
1 & 0\\
2 & 4
\end{pmatrix}
\begin{pmatrix}
H\\
K\\
\end{pmatrix},
\end{equation*}

\begin{equation*}
\begin{pmatrix}
H\\
K\\
\end{pmatrix}_2=
\begin{pmatrix}
0 & 1\\
-4 & -2
\end{pmatrix}
\begin{pmatrix}
H\\
K\\
\end{pmatrix},
\end{equation*}
\begin{equation*}
\begin{pmatrix}
H\\
K\\
\end{pmatrix}_3=
\begin{pmatrix}
-1 & -1\\
2 & -2
\end{pmatrix}
\begin{pmatrix}
H\\
K\\
\end{pmatrix},
\end{equation*} 
as illustrated in Fig.~\ref{fig:overall}(d). The lattice points of the three twin domains [integer $(H,K)_i$ shown as orange, blue, and green dots] are well-separated, except at lattice points of the high temperature phase [integer $(H,K)$ shown as larger black dots]. 


Comparison of single crystal XRD diffraction $[HK2]$ maps at 40~K and 60~K reveal the systematic appearance of CDW peaks below $T_{\rm CDW}$ [Fig.~\ref{fig:overall}(e)]. The CDW peaks are roughly two orders of magnitude weaker than the main Bragg peaks, and the occurrence of CDW peaks at both ${\bf q}_{\rm CDW}$ and $2{\bf q}_{\rm CDW}$ evidence a non-sinusoidal charge density modulation [Fig.~\ref{fig:overall}(f)], consistent with previous work \cite{Liu2024}. The CDW peaks are also systematically observed in the $[H0L]$ plane across numerous Brillouin zones [Fig.~\ref{fig:overall}(g)], and the absence of rod-like diffuse scattering along $L$ indicates the CDW is three-dimensional, and without significant stacking faults.

\begin{figure*}[t]
    \centering
    \includegraphics[width=1\linewidth]{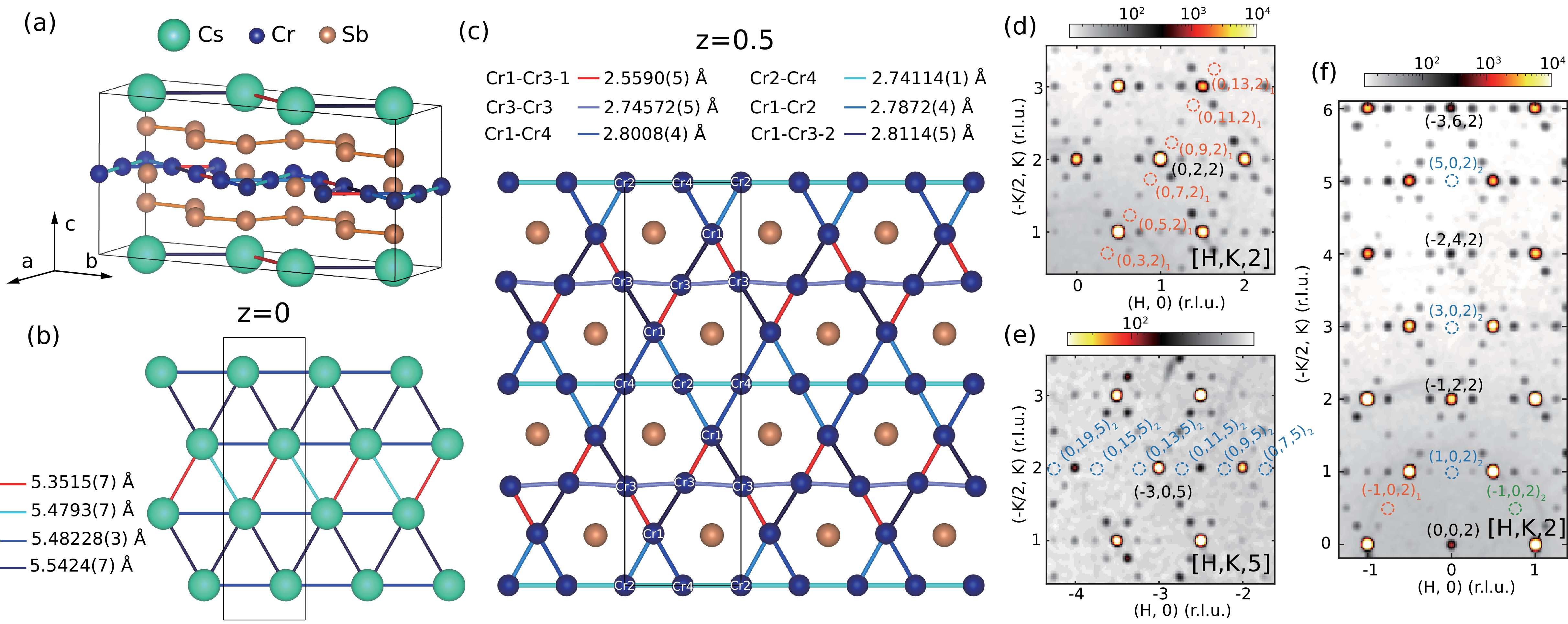}
    \caption{{\bf CDW structure of {\CCS}.} (a) Crystal structure of {\CCS} in its CDW phase. The Cs-Cs and Cr-Cr bonds are distinguished by color based on their lengths. (b) The CDW structure in the $z=0$ plane. (c) The CDW structure in the $z=0.5$ plane. (d) $[HK2]$ and (e) $[HK5]$ maps, showing the systematically forbidden $(0K2)_{1}$ and $(0K5)_{2}$ superstructure peaks with odd $K$, represented by dashed circles. (f) Odd-$H$ $(H02)_{2}$ peaks in the $[HK2]$ plane are shown by dashed circles. Indexing of the high temperature $P6/mmm$ phase Bragg peaks are in black, and CDW peaks due to twin $i$ ($i=1,2,3$) of the $Pbam$ phase are differentiated by color and subscripts $i$.}
    \label{fig:struct}
\end{figure*}

In previous XRD measurements of {\CCS}, a monoclinic distortion was proposed for the CDW state, based on the splitting of Bragg peaks along ${\bf Q}=(-1,5,L)$ and $(-1,5.25,L)$ in XRD \cite{Liu2024}. In Fig.~\ref{fig:overall}(g), similar cuts are shown for our present XRD data, which do not show such a splitting, suggesting that a monoclinic distortion is either absent or much smaller than what previous measurements indicated. As our {\CCS} sample shows CDW peaks similar to previous measurements, this means that a monoclinic distortion is not essential for the $4\times1$ CDW state.

The high-quality of our XRD data allows for a solution of the CDW structure for {\CCS}, which yields an orthorhombic $Pbam$ (space group No.~55) structure, as detailed in Table~\ref{struct_table}. The CDW structure is shown in Fig.~\ref{fig:struct}(a), with Cr-Cr and Cs-Cs bonds of different lengths distinguished by color. The most prominent feature of the CDW phase is the formation of Cr-Cr and Cs-Cs dimers (red bonds), with bonds lengths that are respectively 7.8~\% and 3.0~\% shorter than the average of other Cr-Cr and Cs-Cs bonds. The $z=0$ and $z=0.5$ planes of the CDW structure are further highlighted in Figs.~\ref{fig:struct}(b) and (c). As can be seen, the ordering of the Cr-Cr dimers explicitly breaks sixfold rotational symmetry of the kagome lattice, and is likely the origin of electronic nematicity reported in ultrafast optical spectroscopy measurements \cite{Liu2024_2}.

For the refined CDW structure, systematic absences are expected at odd-$H$  $(H0L)_i$ and odd-$K$ $(0KL)_i$ positions. These peaks are indeed systematically absent in our XRD data, as shown in the $[HK2]$ and $[HK5]$ maps [Figs.~\ref{fig:struct}(d)-(f)], with the forbidden peaks marked by dashed circles. 



\begin{figure}[t]
    \centering
    \includegraphics[width=1\linewidth]{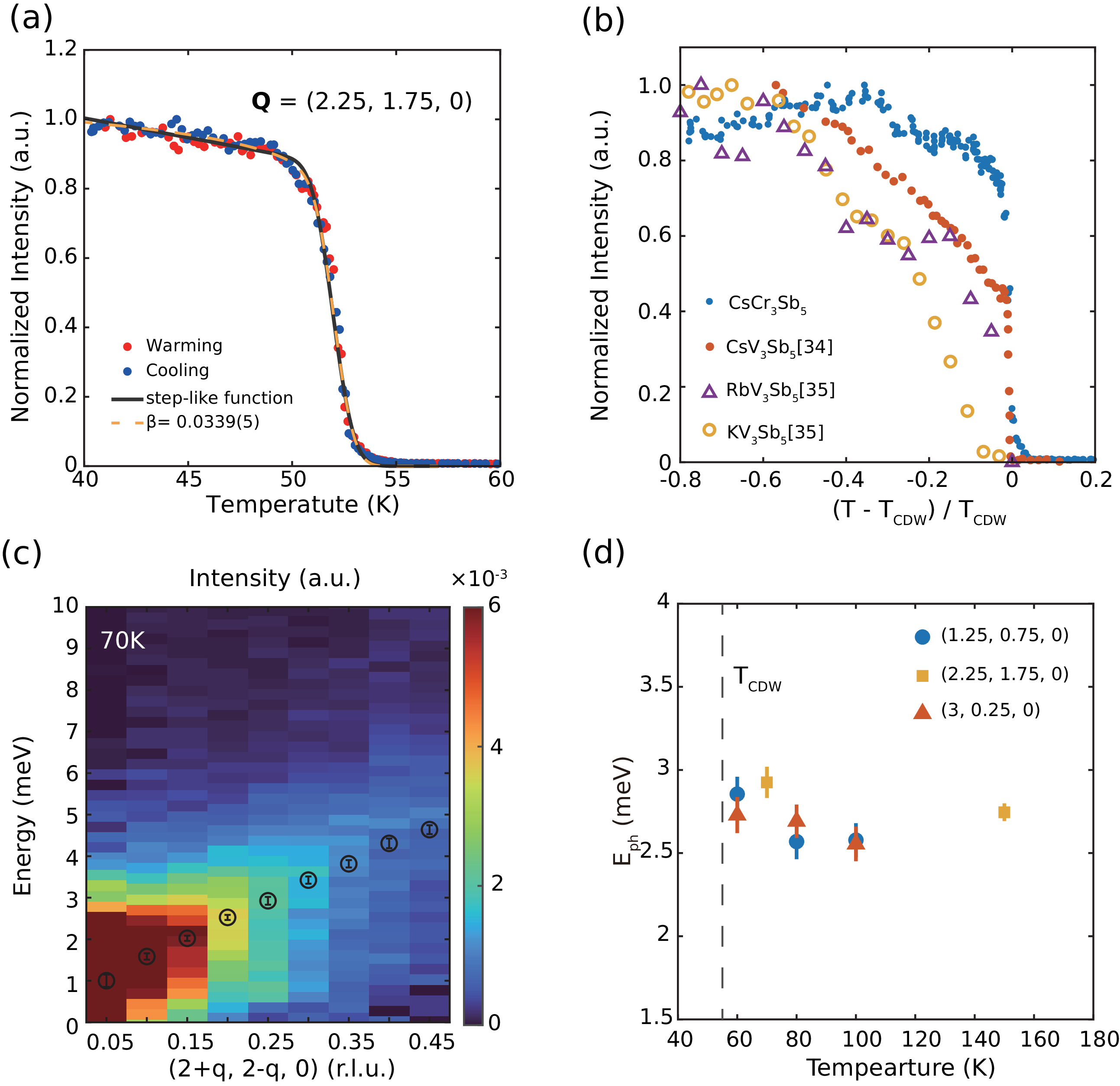}
    \caption{{\bf CDW temperature dependence and phonon measurements.} (a) Temperature dependence of the ${\bf Q}=(2.25,1.75,0)$ CDW peak. (b) Comparison of the CDW peak temperature dependence in {\CCS} and {\AVS}. The data for CsV$_3$Sb$_5$ is from Ref.~\cite{Xiao2023b}, and the data for RbV$_3$Sb$_5$ and KV$_3$Sb$_5$ are from Ref.~\cite{scagnoli2024}. The intensities at low temperatures are normalized to unity. (c) Phonon measurements along $(2+q,2-q,0)$ at 70~K. The open circles are fitted phonon energies $E_{\rm ph}$. (d) Temperature dependence of the acoustic mode at $E_{\rm ph}$ probed in several Brillouin zones.}
    \label{fig:IXS}
\end{figure}

\subsection{First order CDW transition and phonon measurements}
An absence of thermal hysteresis in resistivity and magnetization across $T_{\rm CDW}$ suggests the coupled CDW and magnetic transitions in {\CCS} to be either second order or weakly first order \cite{Liu2024}. Using the elastic channel of inelastic X-ray scattering, we probed the intensity of the CDW peak $\mathbf{Q}=(2.25,1.75,0)$, which shows saturation within a few Kelvin of its onset [Fig.~\ref{fig:IXS}(a)], directly evidencing a first order transition. A negligible thermal hysteresis is observed between warming and cooling, consistent with physical property measurements \cite{Liu2024}. A step-like transition with a Gaussian distribution of transition temperatures gives an excellent fit of the measured CDW intensity, yielding $T_{\rm CDW}\approx52$~K and a width of $0.8$~K. A fit to $I\propto(T_{\rm CDW}-T)^{2\beta}$ with a Gaussian distribution of transitions temperatures leads to $\beta\approx0.03$, where the small exponent also excludes the transition being second order.

Temperature dependence of the normalized CDW peak intensities are compared between {\CCS} and {\AVS} compounds in Fig.~\ref{fig:IXS}(b). The CDW transitions are first order in CsV$_3$Sb$_5$ and RbV$_3$Sb$_5$ , and becomes second order in KV$_3$Sb$_5$ \cite{Song2022,Zhang2024,scagnoli2024}. In terms of the discontinuity in the CDW peak intensity, the discontinuity weakens from CsV$_3$Sb$_5$ to RbV$_3$Sb$_5$, and disappears in KV$_3$Sb$_5$. In {\CCS}, the CDW transition is sharper (with larger discontinuity) than the {\AVS} compounds, which indicates the transition is more strongly first-order, although still weakly first order such that a significant thermal hysteresis is not observed. 

No signatures of thermal diffuse scattering were observed in the XRD images at 60~K [Figs.~\ref{fig:overall}(e) and (g)], which corresponds to $(T-T_{\rm CDW})/T_{\rm CDW}\approx0.15$, in contrast to CsV$_3$Sb$_5$ that shows diffuse scattering associated with the CDW up to at least $(T-T_{\rm CDW})/T_{\rm CDW}\approx0.21$ \cite{Subires2023}. Recent IXS measurements indicate the diffuse scattering in {\AVS} result from soft phonons \cite{McGuinness_arXiv2025,Wang_arXiv2025}, and the absence of diffuse scattering in {\CCS} suggests that compared to {\AVS}, similar soft phonons are either absent or less prominent in {\CCS}, consistent with its more strongly first order CDW transition. 

An absence of prominent soft phonons is consistent with our IXS measurements, which only shows an acoustic phonon mode along $(2+q,2-q,0)$ at 70~K, without signatures of soft modes at ${\bf q}_{\rm CDW}$ [Fig.~\ref{fig:IXS}(c)]. Phonon measurements were also performed in other Brillouin zones, also without ${\bf q}_{\rm CDW}$ soft phonons, and the detected ${\bf q}_{\rm CDW}$ acoustic phonons are essentially independent of temperature [Fig.~\ref{fig:IXS}(d)]. Raw and additional IXS data are presented in Supplementary Note 1 and Supplementary Figs.~1-3.
A similar absence of soft phonons was also reported in {\CCS} with short-range CDW \cite{Yao2024}.

A divergent $E_{2g}$ elastoresistance above $T_{\rm CDW}$ has been reported for {\CCS} \cite{Liu2024_2}, which suggests strong nematic fluctuations in the high-temperature phase. However, such nematic fluctuations do not lead to anomalies of the in-plane transverse acoustic phonons with the same symmetry [Fig.~\ref{fig:IXS}(c) and (d)], as widely observed in the parent compounds of iron-based superconductors \cite{Niedziela2011,Parshall2015,Li2018}. This suggests a weak coupling between nematicity and the lattice in {\CCS}, which occurs in the iron-based superconductors upon overdoping \cite{Wu2025_2}.


\begin{figure}[t]
    \centering
    \includegraphics[width=1\linewidth]{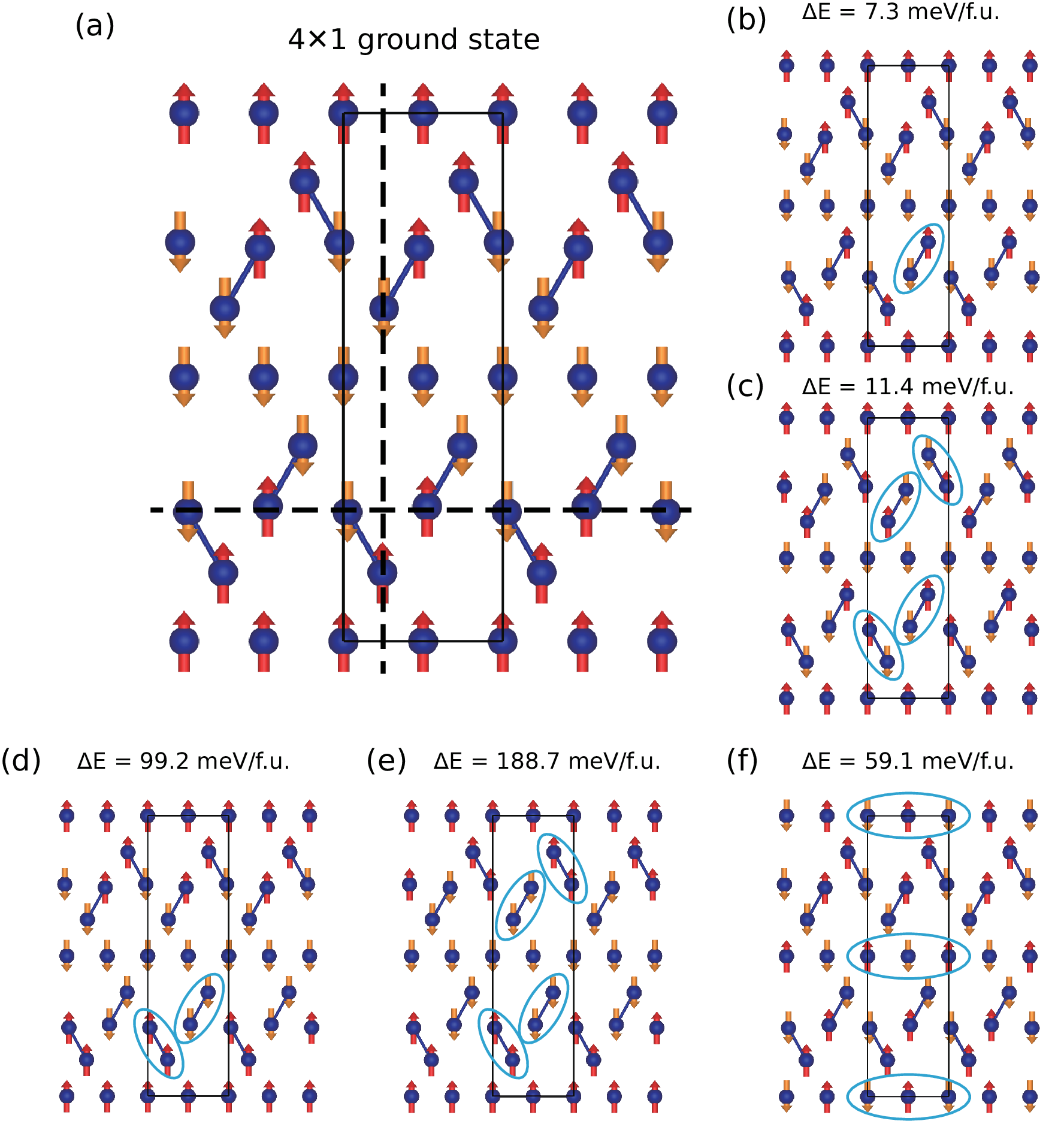}
    \caption{{\bf Magnetic structure and magnetic interactions of the $4\times1$ CDW state.} (a) The calculated ground state magnetic configuration of {\CCS}. The black dashed lines indicate the glide planes.  (b)-(f) Excited configurations with flipped spins (highlighted by circles) relative to the ground state configuration, and associated energy costs per formula unit (f.u.).}
    \label{fig:DFT1}
\end{figure}

\begin{figure}[t]
    \centering
    \includegraphics[width=1\linewidth]{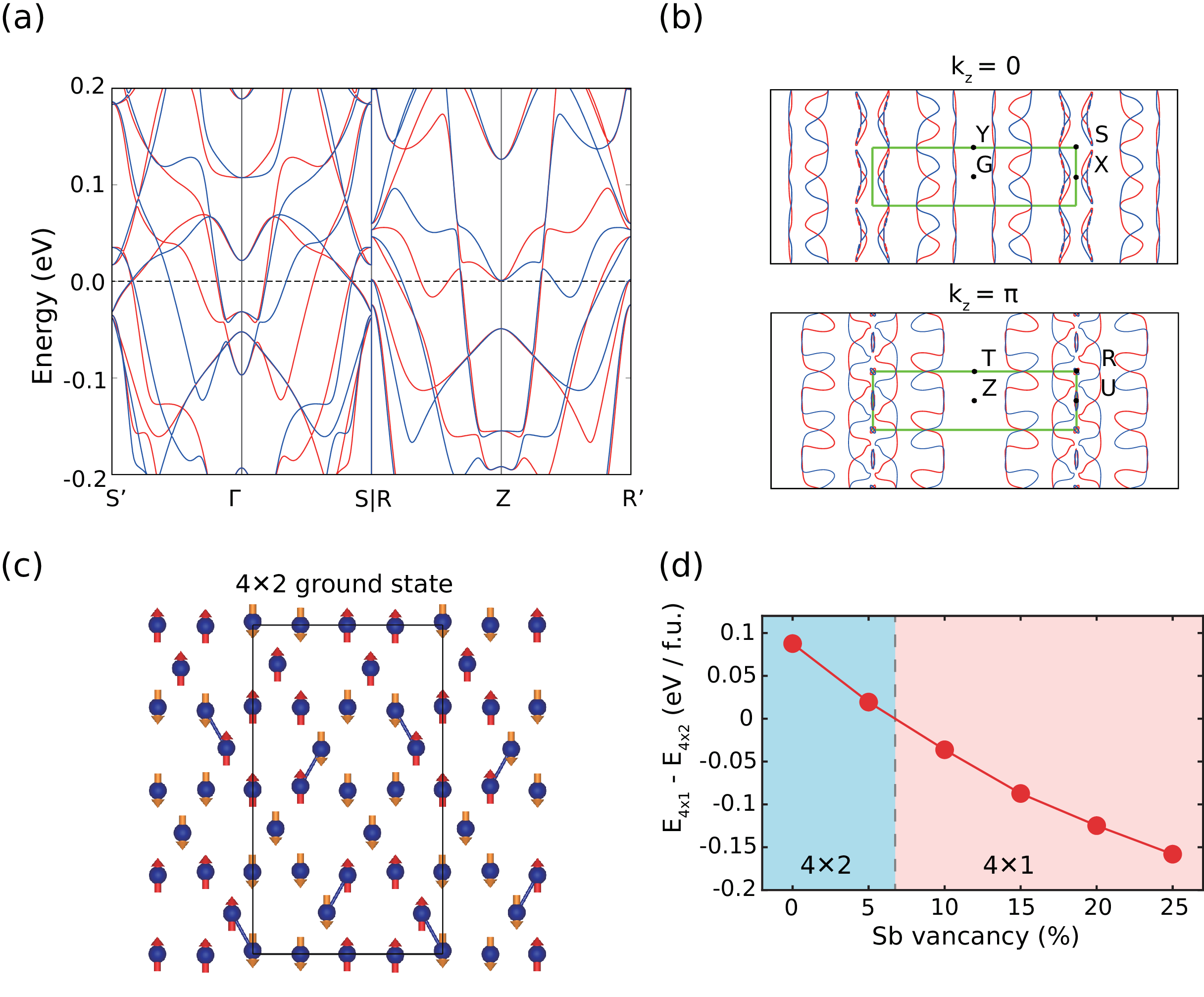}
    \caption{{\bf Electronic structure of the $4\times1$ CDW and a competing $4\times2$ CDW.} (a) The spin-resolved electronic structure of {\CCS}, in its $4\times1$ ground state constrained by the $Pbam$ space group. (b) The corresponding Fermi surfaces of {\CCS}, in the $k_z=0$ and $k_z=\pi$ planes. (c) The $4\times2$ ground state of {\CCS}, with the CDW described by a $Pbam$ structure, also showing Cr dimers \cite{Xu2025}. (d) Energy difference between the $4\times1$ CDW and the $4\times2$ CDW in density functional theory calculations. The $4\times2$ structure has the lowest energy for stoichiometric {\CCS}, the introduction of Sn substituting for Sb lowers the energy of the $4\times1$ CDW relative to the $4\times2$ CDW. Above 7\% of  Sn substitution for Sb, the $4\times1$ CDW becomes lower in energy compared to the $4\times2$ CDW.}
    \label{fig:DFT2}
\end{figure}

\begin{table*}
\caption{Refined CDW structure of {\CCS} at 40~K.}

\begin{ruledtabular}

\begin{tabular}{cccccc}
Chemical formula & Crystal system & Space group & $a$ (\AA) & $b$ (\AA) & $c$ (\AA) \\
\hline
{\CCS}& orthorhombic   & $Pbam$& 5.48228(3) & 18.97249(16) & 9.22336(6)
\end{tabular}
\end{ruledtabular}

\vspace{1em}
\begin{threeparttable}
\begin{ruledtabular}
\begin{tabular}{cccccccccccc}
Atom & $x$ & $y$ & $z$ & Occ. & $U_{\mathrm{eq}}$ &
$U_{11}$ & $U_{22}$ & $U_{33}$ &
$U_{12}$ & $U_{13}$ & $U_{23}$ \\
\hline
Cs1 & 0.26152(2) & 0.37695(2) & 1 & 1 &  5.51(1) & 5.43(2) & 5.20(2) & 5.89(2) & -0.46(2) & 0 & 0 \\
Sb1 & 0.24970(2) & 0.37544(2) & 0.5 & 1 &  3.83(1) & 5.43(2) & 2.17(3) & 6.68(2)  & -0.09(1) & 0 & 0 \\
Sb2 & 0.74966(2) & 0.45708(2) & 0.73616(2) & 1 &  3.75(1) & 3.61(2) & 3.91(2) & 3.74(1) & 0.13(1) & 0.00(1) & -0.10(2) \\
Sb3 & 0.77342(2) & 0.29023(2) & 0.25998(2) & 1 &  3.56(1) & 3.45(2) & 3.55(2) & 3.69(1) & -0.05(1) & -0.10(1) & 0.18(1) \\
Cr1 & 0.74748(4) & 0.37167(2) & 0.5 & 1 &  4.07(4) & 3.62(6) & 3.75(6) & 4.84(11) & 0.73(4) & 0 & 0 \\
Cr2 & 0.5 & 0.5 & 0.5 & 1 &  3.87(3) & 3.69(6) & 2.91(7) & 4.99(9) & 0.12(5) & 0 & 0 \\
Cr3 & 0.51822(4) & 2541.8(2) & 0.5 & 1 &  3.84(2) & 3.26(4) & 3.65(5) & 4.61(6) & -0.13(4) & 0 & 0 \\
Cr4 & 1 & 0.5 & 0.5 & 1 &  4.08(3) & 4.54(7) & 3.12(7) & 4.59(9) & 0.06(6) & 0 & 0 \\
\end{tabular}
\end{ruledtabular}
\begin{tablenotes}
\footnotesize
\item $U_{\mathrm{eq}}$ is one third of the trace of the orthogonalized $U_{ij}$ tensor. Values of $U_{\mathrm{eq}}$ and $U_{ij}$ are in units of $0.001$~{\AA}$^2$.
\end{tablenotes}
\end{threeparttable}
\label{struct_table}
\end{table*}

\subsection{Antiferromagnetic dimers in an altermagnetic ground state}
Based on the results in Ref.~\cite{Xu2025}, we performed a high-throughput search for all possible collinear in-plane AFM magnetic configurations compatible with a $4\times1$ CDW supercell with space group $Pbam$, and then relaxed the crystal structures and computed their associated energies. The resulting state with the lowest energy is shown in Fig.~\ref{fig:DFT1}(a), which shows Cr-Cr dimers separated by Cr chains, in remarkable agreement with the experimentally determined CDW structure in Fig.~\ref{fig:struct}(a). The calculated ground state CDW structure also exhibits Cs-Cs dimers [Supplementary Fig.~4], 
consistent with experimental findings [Fig.~\ref{fig:struct}(b)]. Other $4\times1$ $Pbam$ CDW structures with higher energies, and their associated magnetic structures, are presented in Supplementary Note 2 and Supplementary Fig.~5. 

As shown in Fig.~\ref{fig:DFT1}, the dimers in the calculated CDW ground state are AFM, while the Cr chains are FM. To probe the magnetic interactions between Cr atoms, we examined how flipping certain spins affects the energy of the ground state magnetic configuration. By flipping all the spins on Cr dimers [Fig.~\ref{fig:DFT1}(c)], we find that the energy is only increased by 11.4~meV/f.u. For this excited state, all the dimers remain AFM and the Cr chains remain FM, and the relative spin orientations between the AFM dimers are also maintained. Thus, the increased energy results from the couplings between Cr chains and the Cr dimers, which are rather weak. 
To examine the magnetic interactions between AFM dimers, we flipped spins on one (out of four in the unit cell) of the AFM dimers [Fig.~\ref{fig:DFT1}(b)], and found that its energy increased only by $\sim7.3$~meV/f.u., which suggests that the coupling between AFM dimers is also rather weak.

The magnetic interactions on the Cr chains are considered by examining a configuration with AFM chains, which leads to an increase of 59.1~meV/f.u. ($\sim15$~meV/bond), suggesting that the coupling along the Cr chain is larger than the inter-dimer and dimer-chain couplings. Compared to the relatively weak couplings discussed above, the dominant magnetic interaction in {\CCS} is the AFM coupling between spins within the dimers, evidenced by the significant energy costs of configurations with 2 or 4 FM dimers [Figs.~\ref{fig:DFT1}(d) and (e)], which suggest $\approx50$~meV/dimer when the dimer is changed from AFM to FM. Furthermore, our calculations indicate the inter-layer couplings are very weak, such that AFM and FM inter-layer couplings cannot be differentiated within the precision of first-principles calculations. 

The ground state magnetic configuration for the $4\times1$ CDW has zero net magnetization, and because its two spin sub-lattices are related by a $\{M_{yz}|(0,\frac{1}{2},0)\}$ or a $\{M_{xz}|(\frac{1}{2},0,0)\}$  glide mirror plane [dashed lines in Fig.~\ref{fig:DFT1}(a)] rather than translation or inversion, it is in fact an altermagnetic state \cite{Smejkal2022,Cheong2025,Liu2025,Song2025} in the two-dimensional limit. In three dimensions, a FM interlayer coupling between spins will lead to an altermagnetic state in the bulk, whereas an AFM coupling will lead to the cancellation of spin polarization between adjacent layers, resulting instead in a hidden altermagnetic state \cite{Yang2025}. Assuming a FM interlayer coupling in the CDW state, Fig.~\ref{fig:DFT2}(a) shows the spin polarized band structure of {\CCS}, directly demonstrating its altermagnetic nature. The corresponding Fermi surface is shown in Fig.~\ref{fig:DFT2}(b), further evidencing spin-polarized bands. 

When the CDW structure is not constrained to the experimentally determined $4\times1$ cell, a $4\times2$ CDW structure is calculated to have a lower energy [Fig.~\ref{fig:DFT2}(c)] \cite{Xu2025}. This $4\times2$ CDW is also altermagnetic with the same crystal symmetry and consists of AFM Cr-Cr dimers \cite{Xu2025}, similar to the experimentally determined $4\times1$ CDW [Fig.~\ref{fig:struct}(a)]. This shows that while distinct phases compete to be the ground state of {\CCS}, the presence of AFM dimers is a robust feature of these phases. 
At the level of density functional theory, the $4\times1$ CDW can be stabilized as the ground state by introducing a small amount of Sn substituting for Sb (7\%), modeled via the virtual crystal approximation [Fig.~\ref{fig:DFT2}(d)]. This suggests that the $4\times2$ and $4\times1$ CDWs are close in energy, with the latter possibly stabilized by deviations from an ideal stoichiometry. Alternatively, strong electronic correlations in {\CCS} beyond density functional theory may tilt the balance in favor of the $4\times1$ CDW, and render the $4\times2$ state less competitive. 

\section{Discussion and Conclusion}

Our findings reveal that the CDW structure of {\CCS} is very different from those in {\AVS}, ScV$_6$Sn$_6$, and LuNb$_6$Sn$_6$. In {\AVS}, nonmagnetic V atoms on the kagome lattice form star of David or inverse star-of-David structures that result in a $2\times2$ CDW in the $ab$-plane \cite{Wilson2024}. In ScV$_6$Sn$_6$ and LuNb$_6$Sn$_6$, the $3\times3\times3$ CDWs are dominated by the motion of Sn-Sc(Lu)-Sn trimers along $c$-axis, rather than in-plane distortions of the V or Nb kagome lattice \cite{Arachchige2022,Ortiz2025}. The in-plane modulations of these CDW structures retain the sixfold rotational symmetries of their high temperature phases, which maybe broken by the stacking of distorted planes along the $c$-axis. On the other hand, the CDW structure of {\CCS} is inherently strongly anisotropic, with Cr-Cr dimers separated by Cr chains that run along the $a$-axis of the $4\times1$ cell [Fig.~\ref{fig:struct}(a)], which in combination with three possible twin domains [Fig.~\ref{fig:overall}(c)], naturally account for its three-state Potts nematicity \cite{Liu2024_2}. 

As the ideal kagome lattice with AFM nearest-neighbor interactions is highly frustrated, CDW formation provides a natural way to relieve this mangetic frustration and facilitate long-range magnetic order. Therefore, the CDW of {\CCS} is likely driven by the spin Jahn-Teller effect, which saves magnetic energy at the cost of elastic energy. For comparison, the CDWs in {\AVS} are driven by the nesting between van Hove singularities close to the Fermi level \cite{Tan2021,Park2021,Lin2021,Denner2021} or electron-phonon coupling \cite{Luo2022,Liu2022,He2024,You2025,Wang_arXiv2025,McGuinness_arXiv2025}, whereas the CDWs in ScV$_6$Sn$_6$ and LuNb$_6$Sn$_6$ are driven by electron-phonon coupling \cite{Cao2023,Korshunov2023,Hu2024_ARPES}, potentially also involving nesting effects \cite{Yang2025_166}. 

Under pressure, the intertwined CDW and magnetic transitions of {\CCS} split and are continuously suppressed \cite{Liu2024}, similar to magnetic and nematic orders in the iron-based superconductors \cite{Dai2015,Fernandes2019,Bhmer2022}. Our findings show that the AFM coupling within Cr-Cr dimers is the dominant energy scale in the CDW state of {\CCS}, with the associated fluctuations likely persisting beyond the the suppression of the CDW, and contribute to the non-Fermi-liquid state from which a dome of SC emerges. Resonant inelastic X-ray scattering detected two spin excitation modes in {\CCS} \cite{Wang2025}, and the dispersionless nature of these modes suggest they are associated with local entities, for which the AFM dimers uncovered in this work are natural candidates. In such a scenario, the persistence of such local excitations above $T_{\rm CDW}$ \cite{Wang2025} suggests that dynamic variants of ordered dimers in the CDW state survive even when the CDW is destabilized.

SC at the verge of dimer formation has been reported in IrTe$_2$ \cite{Yang2012,Pyon2012}, AuTe$_2$ \cite{Kitagawa2013,Kudo2013}, and CuIr$_2$S$_4$ \cite{Chen2026}. However, in these cases the dimers are nonmagnetic, the dimer state is suppressed via first-order-like quantum phase transitions, and the associated SC emerge from Fermi liquid normal states. {\CCS} is thus unique compared to these systems, as its dimers are magnetic with a strong AFM coupling, the intertwined CDW and magnetic orders are continuously suppressed towards a putative QCP, and the SC emerges from a non-Fermi-liquid normal state.  

While our DFT calculations suggest ordered AFM dimers that lead to an overall altermagnetic state, it takes very little energy to flip both spins of a given dimer. This raises the interesting possibility that the dimers are in fact spin singlets (rather than AFM-ordered), which could become mobile as the CDW is suppressed under pressure.  Such a scenario is akin to the long-sought goal of obtaining a superconductor by doping a resonating valence bond state or a valence bond solid \cite{Anderson1973,Anderson1987,Shimizu2007}, where mobile spin singlets become Cooper pairs. While this possibility is tantalizing, we note the superconducting coherence length of {\CCS} from upper critical field measurements is around 5~nm \cite{Liu2024}, much larger than the dimers observed in the parent phase.


In summary, we showed that the CDW of the correlated kagome metal {\CCS } is characterized by a $Pbam$ structure consisting of AFM Cr-Cr dimers and FM chains, explicitly breaking the sixfold rotation symmetry of the high temperature phase. The CDW transition in {\CCS} is found to be more strongly first order than its vanadium-based sister compounds, likely accounting for the absence of diffuse scattering or prominent soft phonons above $T_{\rm CDW}$. Constrained by the experimentally determined structure, we obtained the magnetic structure of {\CCS} via a systematic search of collinear magnetic phases, and showed the dominant magnetic interactions is AFM and within the Cr dimers. The CDW and magnetic structures determined in this work form the basis for further understanding {\CCS}, in which kagome physics, unconventional superconductivity, and strong electronic correlations intersect.

\section{Acknowledgments}
This work was supported by the National Key R\&D Program of China (No.~2022YFA1402200), the National Natural Science Foundation of China (No.~12350710785, No.~12274363, No.~12304175, and No.~12474132), the Beijing National Laboratory for Condensed Matter Physics (Grant No.~2023BNLCMPKF019), and the Fundamental Research Funds for the Central Universities (Grant No. 226-2024-00068). The synchrotron single-crystal X-ray diffraction and inelastic X-ray scattering experiments were carried out at BL02B1 and BL35XU of SPring-8 with the approval of the Japan Synchrotron Radiation Research Institute (JASRI) (Proposals No.~2025A1835 and No.~2024B1724).

\section{Methods}
\subsection{Experimental details}
High-quality {\CCS} single crystals were grown using the self-flux method \cite{Liu2024}. Single crystal XRD measurements were performed at the BL02B1 beamline at SPring-8 with 39.9~keV X-rays. Determination of the sample orientation and reciprocal space reconstruction were performed using the \texttt{CrysAlis} software package. Structural refinement was performed using the \texttt{Olex2} software.

IXS measurements were carried out at the BL35XU beamline at SPring-8 \cite{Baron2000}. The measurements were formed in the transmission geometry with 21.7 keV X-rays. A {\CCS} sample $\sim2$~$\mu$m in thickness was mounted on a copper sample holder with GE varnish. 50 ~mm$\times$50~mm slits were used in IXS measurements to maximize the scattering intensity. The instrumental resolution $R(E)$ was obtained by measuring a piece of tempax glass, normalized to unit area, and parametrized using a pseudo-Voigt function. All measured scattering intensities were normalized by a monitor before the sample.

The measured IXS spectra $I(E)$ are modeled by a small constant term $b$, a resolution-limited elastic peak $c$, and the scattering function $S(E)$ convolved with the instrumental resolution $R(E)$:
\begin{equation*}
	I(E) = b+cR(E)+\int_{-\infty}^{\infty} [S(E -E')]R(E') dE'.
    \label{eq:Intensity}
\end{equation*}

The scattering function of a phonon mode is modeled as a generalized damped harmonic oscillator (DHO): 
\begin{equation*}
S_{\rm DHO}(E)=\frac{1}{1-\exp(-\frac{E}{k_{\rm B}T})}\frac{A_{\rm ph}}{E_0}\frac{2\gamma E E_0}{\pi[(E^2-E_0^2)^2+(E\gamma)^2]},
\label{eq:DHO}
\end{equation*}
where $E_0$ is the undamped phonon energy, $A_{\rm ph}$ is the phonon intensity factor, and $\gamma$ is the damping rate. $\gamma$ renormalizes the phonon energy from $E_0$ to $E_{\rm ph}=\sqrt{E_0^2-\gamma^2/4}$. For the phonons probed in this work, $\gamma$ is relatively small and thus $E_0\approx E_{\rm ph}$. 

\subsection{First-principles calculations}
The calculations were performed using density functional theory (DFT) with the \texttt{VASP} package \cite{PhysRevB.47.558,PhysRevB.59.1758}. The energy cutoff of the plane-wave basis was chosen to be 450 eV and a $\Gamma$-centered 9$\times$18$\times$6 $\mathbf{k}$-point mesh was employed for the 4$\times$1 CDW structure. 
In our calculations, the PBEsol approximation\cite{PBEsol} was used and spin-orbit coupling (SOC) was not included. For all the 4$\times$1 $Pbam$ CDW structures generated from our high-throughput search, the lattice constants were constrained to experimentally determined values at 40~K, and the atomic coordinates were fully relaxed until the force on each atom is less than 1~meV/{\AA} and the internal stress is less than 0.1~kbar for . To model Sn substituint for Sn, we employed the virtual crystal approximation (VCA) as implemented in \texttt{VASP}. In this approach, the pseudopotential for the mixed Sb/Sn site is constructed through a linear combination of the atomic potentials of Sb and Sn.

\bibliography{CsCr3Sb5.bib}

\end{document}